\documentclass[a4paper,11pt]{article}
\pdfoutput=1 

\usepackage{jcappub} 

\usepackage[T1]{fontenc} 

\newcommand{\dd}{\mathrm{d}}

\title{\boldmath Modified Magnetohydrodynamics Around the Electroweak Transition}

\author{Petar Pavlovi\'c,}
\author{Natacha Leite}
\author{and G\"unter Sigl}

\affiliation{II. Institut f\"ur Theoretische Physik, University of Hamburg\\
Luruper Chaussee, 149, 22761 Hamburg, Germany}

\emailAdd{petar.pavlovic@desy.de}
\emailAdd{natacha.leite@desy.de}
\emailAdd{guenter.sigl@desy.de}

\abstract{
We analyse solutions of the MHD equations around the electroweak transition taking into account the effects of the chiral anomaly. It is shown that a transition that is not of the first order has direct consequences on the evolution of the asymmetry between left- and right-handed leptons. Assuming an initial chiral asymmetry in the symmetric phase at temperatures higher than the transition temperature, as well as the existence of magnetic fields, it is demonstrated that the asymmetry typically grows with time, until it undergoes a fast decrease at the transition, and then eventually gets damped at lower temperatures in the broken phase. We argue that it is unlikely to have any significant magnetic field amplification as a consequence of the electroweak transition in the Standard model, even when the chiral anomaly is introduced. The presence of a chiral asymmetry between left- and right-handed charge carriers naturally leads to the creation of helical magnetic fields from non-helical fields and this can have consequences on their subsequent evolution. Similarly, an initially vanishing chiral asymmetry is naturally created in the presence of a helical magnetic field.}

\begin{document}
\maketitle
\flushbottom

\section{Introduction}
\label{sec:intro}
Experimental results confirm that magnetic fields are present on all scales of the observable universe: from stars and planets, to
clusters of galaxies and even to the voids of the intergalactic medium (\cite{Beck, Bernet, Vogt, Neronov, J.P. Vallee, R.Durrer}). Magnetic field measurements, using Faraday rotation and Zeeman splitting  methods, suggest that the typical strength of magnetic fields in galaxy clusters 
is of the order of $10^{-6} $ Gauss with correlation scales of the order of tens of kpc, and that even the inter-galactic voids have magnetic
fields of the order of $10^{-16}$ Gauss \cite{Neronov}. The question of the origin of seed fields, which could then be amplified to the presently observed ones via the galactic dynamo mechanism \cite{Parker}, is still unsolved. There are two main approaches
when dealing with it: assuming that these fields can be created by charge separation during galaxy formation (i.e. that they have an
astrophysical origin) \cite{Subramanian} or that they essentially have a cosmological origin and are related to processes in the early universe. Of course, assuming a cosmological origin for the magnetic fields
requires further elaboration of seed field production mechanisms in the cosmological context, an appealing framework for this being given by 
distinctive periods in the evolution of the universe -- such as inflation \cite{Turner, Bamba, Giovannini} or phase transitions 
\cite{Joyce, Enqvist, Sigl}. Further amplification of seed magnetic fields can be produced during galaxy cluster formation in processes including turbulence created by mergers and buoyancy of the cluster medium \cite{beresnyak}.

In any case, assuming that 
magnetic fields were present in the early universe seems natural and there are no compelling reasons against it -- regardless of their 
potential later role in producing the present galactic fields. Therefore we will also assume their presence at the period of the electroweak 
 transition. Moreover, we will also assume that these fields were possibly helical, i.e. had a non-vanishing helicity density, 
$ h=V^{-1}\int  \textbf{A} \cdot \textbf{B} \, \dd^3 \mathbf{r} $. The production of such fields was previously proposed in the context of inflation,
as well as for the electroweak and QCD phase transitions \cite{Anber,Cornwall, Copi}. Nevertheless, we will show that assuming the existence of helical fields is not crucial for our analysis and that helical fields can be created in the context of modified
magnetohydrodynamics even if no initial helicity was present.
The early universe was characterized by very high conductivity, so it can be described in the usual magnetohydrodynamic picture
\cite{Baym, Giovannini2}. It was also recently argued that for temperatures higher than 10 MeV the typical magnetohydrodynamic
(MHD) system of Maxwell and Navier-Stokes equations should be extended to take into account the effect of chiral anomaly \cite{Boyarsky,Vachaspati}.
 In this work we will investigate the influence of the electroweak transition 
in the Standard Model on the evolution of a chiral asymmetry and, therefore, on the solutions of the modified MHD equations. 

In the remainder of this section we revisit the concept of the chiral anomaly, its application to the early universe and basic assumptions taken throughout. In \S\ref{sec:MHD} we define the MHD equations before and after electroweak symmetry breaking and estimate the chirality flipping rates in each regime. Afterwards, in \S\ref{sec:sol_analytical} we study these equations analytically and predict the behavior of relevant variables and in \S\ref{sec:sol_numerical} we show the results of studying them numerically. Finally we summarize and conclude in \S\ref{sec:conclusions}.

\subsection{Chiral anomaly in the early universe}
At the high temperatures of the early universe all processes involving electron mass are suppressed. In this regime one can 
therefore approximately neglect the effects of the difference between helicity and chirality operators, and introduce the  
density of left/right chiral electrons, $n_{L,R}=(2V)^{-1}\int \psi^{\dagger} (1 \pm \gamma_{5}) \psi \dd^3x$, as well 
as the respective chemical potentials, $\mu_{L,R}= 6 n_{L,R}/T^{2}$, and their difference, $\mu_{5}\equiv (\mu_{L} - \mu_{R})/2$. Electron chiral number 
densities are approximately conserved in the absence
of reactions that flip the chirality of the interacting particles, therefore we can treat them as conserved for temperatures where
chirality flipping rates are out of equilibrium -- i.e. smaller than the Hubble rate, $H(T)$ -- or negligible. On the other hand, for lower
temperatures chirality flipping processes need to be taken into account and they will tend to drive $\mu_{5}$ to zero. In the context
of the electroweak transition it is precisely the difference between the relevant flipping processes in the symmetric and broken phase that introduces
the change in the evolution of $\mu_{5}$ and therefore -- as we will show -- in the solutions of the MHD equations, regardless of the
order of the phase transition.

Even if the chirality flips $\Gamma_f$ could be neglected, in the presence of external 
electromagnetic fields with non-vanishing helicity, $\mu_{5}$ will not stay constant because of a quantum effect 
called chiral anomaly \cite{Kharzeev, Vilenkin, Vilenkin2}. This effect comes from the fact that in the presence of $\mu_{5}$, the contributions 
of left- and right-handed fermions do not cancel each other completely, so that as a
consequence, the axial anomaly creates an electrical current. Moreover, and what is essential for its potential role in cosmology,
this quantum effect can effectively operate on macroscopic scales. This can be clearly seen if we consider the example of a Dirac
sea of massless fermions: when no external fields are applied chirality is conserved and there are two different
Fermi surfaces for left- and right-handed fermions. If we now switch on an electric and magnetic field $\textbf{E}$ and $\textbf{B}$ parallel to each other, a change of chirality that induces an electrical current occurs.
The external magnetic field will tend to orient the spins of negatively charged fermions in an anti-parallel direction with respect to the field (and parallel for positively charged fermions). This comes as a consequence of the orientation that corresponds to the
minimal level of interaction energy $W= -{\boldsymbol \mu_m} \cdot\bf{ B}$, where ${\boldsymbol \mu_m} = gq{\bf s}/(2m_e)$ is the electron ($q=-e$) or positron ($q=e$) spin magnetic moment, with $g$ the spin factor.  At the same time, these fermions will
experience a force $q \mathbf{E}$, and will therefore tend to have a positive projection 
of spin on momentum -- i.e. right-handed helicity for $q=-e$ (and reverse for positively charged fermions). As a consequence, after a time $t$ the momentum of left-handed fermions will be decreased by $p_{L}=-e E t$ and the momentum of right-handed ones will increase by the same amount.
Therefore, the density of states along the direction of the electric and magnetic fields (which we choose to lie along the $z$ axis) will be $\dd N_{R}/\dd z=p_{R}/(2 \pi)$ \cite{Kharzeev}. On the other 
hand, in the transverse directions fermions populate Landau levels in the magnetic field with density
$\dd^{2}N_{R}/\dd x \dd y=eB/(2 \pi)$. Taking into account this increase and the respective decrease in the number of left-handed fermions one has
\begin{equation}
 \frac{\dd(n_{L}-n_{R})}{\dd t}=- \frac{e^2}{2 \pi^2 V} \int \textbf{E} \cdot \textbf{B}\, \dd^{3}\mathbf{r	}.
\end{equation}
Taking into account the chirality flipping rates $\Gamma_{f}$ discussed above, we get
\begin{equation}
\label{mu5}
\frac{\dd \mu_{5}}{\dd t}=\frac{3e^2}{4 \pi^2T^2}\frac{\dd h}{\dd t} - \Gamma_{f}\mu_{5}, 
\end{equation}
where we have expressed the previous integral in terms of helicity density.

Considering the change of energy related to this chirality flow, it can be shown that it corresponds to the electrical current
\begin{equation} \label{eq:j5}
 \textbf{j}_5=- \frac{e^{2}}{2 \pi^{2}}\mu_{5} \textbf{B}\,.
\end{equation}
Therefore, in the presence of a non-vanishing $\mu_{5}$, the usual MHD equations -- consisting of Maxwell, Navier-Stokes, and continuity equations in the resistive MHD approximation \cite{Spruit}, \cite{Giovannini2} -- should also include the contribution 
from this effective current, and they read 
\begin{equation}
 \nabla \times \textbf{B} = 4 \pi\left[ \sigma (\textbf{E} + \textbf{v} \times \textbf{B})- \frac{e^{2}}{2 \pi^{2}}\mu_{5} \textbf{B} \right],
\label{Maxwell}
\end{equation}
\begin{equation}
 \partial_t\textbf{B}= - \nabla \times \textbf{E},
\end{equation}
\begin{equation}
 \rho\left[\partial_t\textbf{v} + (\textbf{v} \cdot \nabla) \textbf{v} - \nu \nabla^{2}\textbf{v}\right]= - \nabla p + 4\pi\sigma[ \textbf{E}\times \textbf{B} + (\textbf{v} \times \textbf{B}) \times \textbf{B}],
\end{equation}
\begin{equation}
\partial_t\rho+ \nabla(\rho \cdot \textbf{v})=0,
\label{c}
\end{equation}
where $\sigma$ is the electrical conductivity, $\rho$ the energy density and $\nu$ the kinematic viscosity. Eq. \eqref{mu5} must also be added to this system. In this approximation of Maxwell's equations high conductivity
is assumed, as well as global neutrality of plasma, i.e. $\nabla \cdot \textbf{j}=0$, $\nabla \cdot\textbf{ E}=0$ and the displacement current is neglected. 

We will be focusing on the electroweak transition in the Standard Model, meaning that bubble collision and turbulence will not contribute to the fluid velocity, since this transition is not of the first order. However, primordial density perturbations could be converted into velocity fluctuations \cite{Wagstaff}. These velocity perturbations could in principle be a source of turbulence in the electroweak transition if the Reynolds number are large enough. To estimate the relative importance of the velocity field compared to the chiral instability, one can consider a fluid velocity spectrum of the form $\langle\boldsymbol{\upsilon}^2\rangle = \boldsymbol{\upsilon}_i^2(k/k_i)^n$, with $n$ the power index.
Rewriting it with respect to the length scale $\ell=2\pi/k$ yields the velocity flow $\displaystyle \boldsymbol{\upsilon}_\ell = \sqrt{\langle\boldsymbol{\upsilon}^2\rangle}(\ell/L)^{n/2}$, with $L$ the integral length scale at which most of the power is concentrated.
The Ohmic current can be approximated to $\mathbf{j}_{\rm Ohm} = \sigma ( \boldsymbol{\upsilon} \times \mathbf{B})$ and thus, comparing it with the anomaly current  \eqref{eq:j5}, one finds
\begin{equation} \label{eq:Jratio}
\frac{j_{\rm Ohm}}{|j_5|} \simeq \frac{2\pi^2}{e^2}\sigma \frac{\upsilon_{\rm rms}}{\mu_5 } \left(\frac{\ell}{L}\right)^{n/2},
\end{equation}
where $\displaystyle \upsilon_{\rm rms} \equiv  \sqrt{\langle\boldsymbol{\upsilon}^2\rangle}$.
For an expanding Universe, the integral scale is given by $L=\upsilon_{\rm rms}/H$, since at the largest scales the time scale can at best be taken as the Hubble time. 
Around the electroweak phase transition, the conductivity will approximately be $\sigma/T \simeq 70$, the Hubble parameter $H/T \sim 10^{-17}$ and we consider the relevant scale to be $\ell_5=2\pi/k_5$, where $k_5$ is given below by \eqref{eq:k5}. This results in $j_{\rm Ohm}/|j_5| \sim 2(\pi/e)^{n+2}\sigma H^{n/2} (\upsilon_{\rm rms})^{1-n/2}/|\mu_5|^{1+n/2}$. 
If we now consider an estimated upper limit $\upsilon_{\rm rms}\sim 5\times 10^{-5}$ \cite{Wagstaff} and assuming a Kolmogorov turbulence spectrum with $n=2/3$, the chiral term dominates for $(\mu_5/T) \gtrsim \upsilon_{\rm rms}^{1/2} \sim 10^{-2}$. 
In the remainder we neglect the effects of fluid velocity, leaving the more general question of the interplay between turbulence and chiral asymmetry for later work.

We furthermore assume that fields are slowly varying so that chemical potentials can be treated as
space-independent quantities. The evolution of magnetic fields has to be described in an expanding universe with the Friedmann-Robertson-Walker-Lema\^itre metric
\begin{equation}
 \dd s^{2}=-\dd t^{2} + a(t)^{2}\left[\dd x^{2} + \dd y^{2} + \dd y^{2}\right]\,,
\end{equation}
where the scale factor $a(t)$ has to be a solution of the Friedmann equations. 
It can be shown that MHD equations stay the same as in the flat spacetime if $t$ is replaced by the conformal time $dt=a(t) d\tau$, with $a(t)=T^{-1}$, and all variables replaced by their conformal counterparts: $\textbf{B} \rightarrow a(t)^{2}\textbf{ B}$, $\textbf{E} \rightarrow a(t)^{2} \textbf{E}$, $\sigma \rightarrow a(t)\sigma$, $\mu_5(t)\rightarrow a(t)\mu_5$, $k \rightarrow a(t)k$, $\Gamma \rightarrow a(t) \Gamma$ \cite{Subramanian2, Banarjee}. Since around the electroweak transition the universe is radiation dominated, to convert time into temperature conformal time can be written as $\tau =M_*/T$, with $M_{*}=(90/8 \pi^3 g_{*})^{1/2} M_{Pl}$ the reduced Planck mass, where $g_{*}\approx106.75$ is the number of degrees of freedom around the electroweak transition and $M_{Pl}$ the Planck mass. 
In the next sections all quantities will be written in conformal units.

\section{MHD equations and chirality flipping rate around the electroweak transition} \label{sec:MHD}
\subsection{Magnetohydrodynamics in the electroweak region}

At temperatures higher than the temperature of the electroweak transition, the symmetry group $SU_{L}(2)\otimes U_{Y}(1)$ is restored. For lower temperatures the symmetry is broken down to $U_{EM}(1)$ group, which corresponds to the existence of ordinary electric and magnetic fields. Since electrical fields
decay on a time scale inversely proportional to the high conductivity of the early universe, the only long-range fields that survive are magnetic ones.
Above the transition we are essentially dealing with unified electroweak interactions, but long-range non-Abelian fields decay and the only non-screened 
modes correspond to $U(1)_{Y}$ hypercharge group \cite{Giovannini3}. Unlike classically described magnetic fields, which have a vector-like coupling 
to fermions, hypermagnetic ($H^{Y})$
 and hyperelectric ($E^{Y}$)
 fields introduce the chiral coupling, which is related to the change in fermion number
\begin{equation}
 \partial_{\mu}j^{\mu} \sim \frac{g'^{2}}{2 \pi^{2}}\textbf{B}^{Y}\cdot \textbf{E}^{Y},
\end{equation}
where $g'=e/\cos\theta_{W}$ is the hypercharge coupling constant with $\theta_{W}$ being the Weinberg angle. Note that this
contribution, coming from the Chern-Simons anomaly term in the Standard Model Lagrangian for the hypercharge field, is analogous 
to the previously described term coming from the chiral anomaly contribution when dealing with ordinary magnetic and electric fields. 
If the electroweak plasma is in complete equilibrium it can be described by $n_{f}$ chemical potentials, related to the number
of conserved global charges \cite{Giovannini2}
\begin{equation}
 N_{i}=L_{i} - \frac{B}{n_{f}}\,,
\end{equation}
where $L_{i}$ is the lepton number of the \textit{i}-th generation, $n_{f}$ the number of fermionic generations and $B$ the baryon number, which holds strictly only when there is no neutrino mixing. In the absence of chirality flips, i.e. at higher 
 temperatures where these processes are out of equilibrium, the number of right-handed electrons is perturbatively conserved and 
one can define the corresponding chemical potential $\mu_{R}$. When dealing with lower temperatures, where chirality flips need to 
be taken into account, one can then perturbatively add the rate of chirality flips to the equations. Moreover, even in the absence of chirality flips, the number of right-handed 
electrons is not exactly conserved because of the aforementioned Abelian anomaly \cite{Giovannini3}
\begin{equation}
 \partial_{\mu} j_{R}^{\mu}=-\frac{g'^{2}y_{R}^{2}}{64 \pi^{2}} Y_{\mu \nu} \tilde{Y}^{\mu \nu}\,,
\end{equation}
with $y_{R}=-2$ being the hypercharge of the right-handed  electron, $Y_{\mu \nu}$ the hypercharge field strength and $\tilde{Y}^{\mu \nu}$ its dual, respectively. 
One can then obtain equations analogous to the previous MHD equations after the symmetry breaking (with Navier-Stokes and continuity
equations staying unaltered, which we here omit since we neglect the velocity effects) \cite{Giovannini2, Giovannini3}
\begin{equation}
 \nabla \times\textbf{ B}^{Y}
 = \sigma\textbf{ E}^{Y}- \frac{g'^{2}}{\pi^{2}}\mu_{R} \textbf{B}^{Y}
\label{hmaxwell}
\end{equation}
\begin{equation}
 \partial_\tau\textbf{B}^{Y}
= - \nabla \times \textbf{E}^{Y},
\end{equation}
\begin{equation}
 \nabla \cdot\textbf{ B}^{Y}=0,
\end{equation}
\begin{equation}
 \nabla \cdot \textbf{E}^{Y}=0,
\label{hmaxwell4}
\end{equation}
to which the anomaly equation should also be added
\begin{equation}
 \frac{\dd n_{R}}{\dd \tau}= \frac{g'^{2}}{4 \pi^{2}}\frac{\dd h^{Y}}{\dd \tau} - \Gamma_{s}n_{R},
\end{equation}
where $\Gamma_{s}$ is the chirality flipping rate in the symmetric region, i.e. before eletroweak transition, and we have introduced hyper-helicity, $
H^{Y}$, defined analogously to the ordinary one. The way in which the chemical 
potential depends on the right-handed number density is not trivial and it is related to the number of fermionic generations, Higgs doublets and other features specific to the elementary particles model. In the minimal Standard Model the evolution equation for the chemical potential of right-handed electrons is given by
\cite{Giovannini3}
\begin{equation}
\frac{\dd \mu_{R}}{\dd \tau}=   \frac{g'^{2}}{8 \pi^{2}}\frac{783}{88}\frac{\dd h^{Y}}{\dd \tau} - \Gamma_{s} \mu_{R} . 
\label{muR}
\end{equation}
\\
\subsection{Chirality flips in the electroweak phase}
\begin{figure}[tbp]
  \centering
 \includegraphics[width=0.6\textwidth]{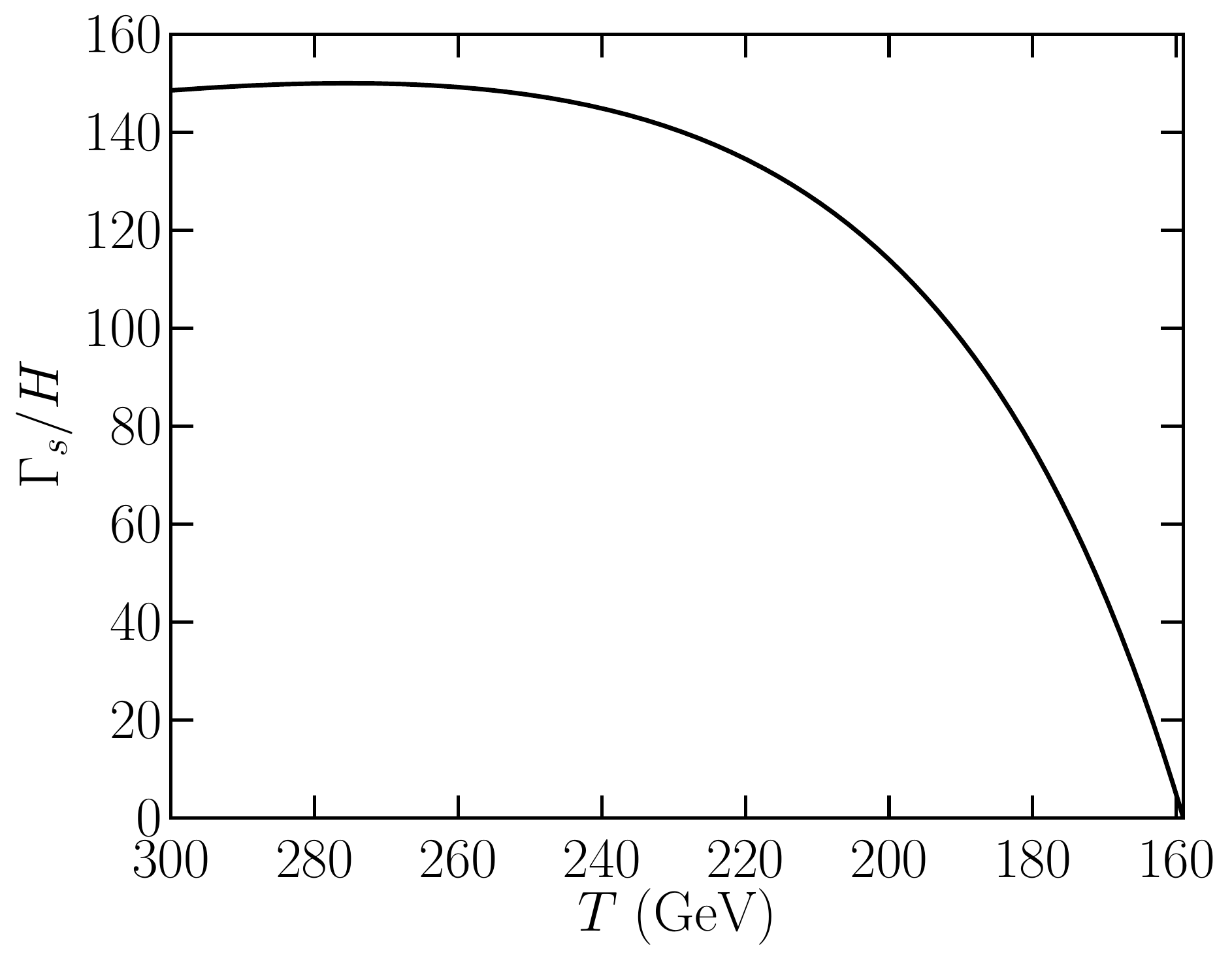}
  \caption{\label{pic:Gamma_s} Chirality flipping rates in the symmetric phase, $\Gamma_{s}$, due to Higgs decays, from \eqref{eq:Gamma_s}, and $t\bar t$-scattering \eqref{eq:ttbar}  normalized to the Hubble rate.}
\end{figure}

Flipping rates before the electroweak transition are determinated by inverse Higgs decays, such as $e_{L} \bar{e}_{R} \leftrightarrow
\varphi^{(0)}$ and $\nu_{e L} \bar{e}_{R} \leftrightarrow \varphi^{(+)}$, with $\varphi^{(+)}$ and $\varphi^{(0)}$ forming the Higgs doublet. 
The rate of inverse Higgs decay per electron is \cite{Cline, Campbell}  
\begin{equation} \label{eq:Gamma_s}
\Gamma_{H} = \frac{\pi}{192 \zeta(3)} h_{e}^{2}  \left(\frac{m(T)}{T}\right)^2,
 \end{equation}
where $m(T)$ is the temperature-dependent effective Higgs mass and $h_{e}$ is the Yukawa coupling for electrons. 
There is also a contribution from scattering processes such as $t_{R} \bar{t}_{L} \leftrightarrow
e_{R} \bar{e}_{L}$. This rate can be estimated from the general expression $\Gamma = n\sigma v$, where $n$ is the particle density, $\sigma$ is the cross-section of the process, computed in Ref. \cite{Protecting} and $v$ is the velocity of the particles involved (which at high temperatures can be taken to be of order unity), allowing us to write the rate as
\begin{equation} \label{eq:ttbar}
\Gamma_{t\bar t} = \frac{(h_t h_{e})^{2}T^2}{8\pi s}\left[\frac{s^2}{(s-m_H^2)^2+\left(\pi h_t^2s/16\right)^2}+2\right],
 \end{equation}
where $h_t$ is the top Yukawa coupling and $s$ is the Mandelstam variable.
Fig.~\ref{pic:Gamma_s} depicts the chirality flipping rates \eqref{eq:Gamma_s} and \eqref{eq:ttbar} before the electroweak symmetry breaking. As can be seen here, and already demonstrated in \cite{Protecting}, inverse Higgs decays are dominating for higher temperatures, but for lower temperatures (corresponding to lower $m_H(T)/T$ values), especially around the transition, they become subdominant compared to the $t\bar t$ processes. When the reaction rate of the latter becomes higher than the first, the rate of Higgs inverse decay is already only of the order of the Hubble rate. Therefore, one could expect that the corresponding flipping rate will be negligible compared to the other terms in the modified MHD equations \eqref{hmaxwell}-\eqref{hmaxwell4}. We have in fact checked that the $t\bar t$ scattering contribution to the flipping rate does not produce an impact on the numerical solutions of the evolution of magnetic fields and asymmetry.

The evolution of the effective Higgs mass will depend 
on the type of phase transition and therefore on the model of elementary particles. We follow the result of lattice simulations \cite{D'Onofrio:2014kta} regarding the order and temperature of the transition in the Standard Model, which is found to be a cross-over. 
On the other hand, in order to obtain an analytical estimate of the dependence of the thermal Higgs mass on temperature, needed to compute the rate of Higgs inverse decays, we use the approximation of 1-loop Higgs potential below. 
The Higgs mass evolution
can be determined from the effective thermal Higgs potential in the high temperature limit to be \cite{Dine, Weinberg} 
\begin{equation} \label{eq:higgspot}
 V(\phi,T) = D(T^{2} - T_{0}^{2})\phi^{2} - ET \phi^{3} + \lambda\frac{\phi^{4}}{4}\,.
\end{equation}
 Parameters $D$, $E$ and $\lambda$ depend on the details of the particle model, where $E$ is of special importance since
it determines the order of the phase transition. 

The Higgs mass is then determined by
$m^{2}=(\dd^{2}V/ \dd  \phi^{2})\mid_{ \phi=v}$, where $v$ is the Higgs expectation value obtained by
$(\dd V/\dd \phi)\mid_{\phi=v}=0$. In the Standard Model, the parameters on Eq. \eqref{eq:higgspot} take the form
\begin{equation}
D=\frac{1}{8v_{0}^{2}}\left(2m_{W}+m_{Z}^{2} + 
2m_{t}^{2}+\frac{m_{H}^{2}}{2}\right)\,, 
\end{equation}
\begin{equation}
E=\frac{1}{4 \pi v_{0}^{3}}(2m_{W}^{3} + m_{Z}^{3})\,,
\end{equation}
\begin{equation}
 T_{0}^{2}=\frac{1}{4D}m_{H}^{2}\,,
\end{equation}
\begin{equation}
\lambda=\left(\frac{m_{H}}{2v_{0}}\right)^{2}\,,
\end{equation}
with $v_{0} \approx 246$ GeV, $m_{H} \approx 125$ GeV and $ 2D \approx 0.38$. From these expressions, 
the Higgs mass changes with temperature as $m(T)^{2}=2D\left(T^{2} - T_{0}^{2}\right)$ and smoothly goes to zero
at $T=T_{0}$. For a first order phase transition, possible for extensions
of the Standard Model, the Higgs mass is changing in the same fashion for $T>T_1$, with
\begin{equation}
 T_{1}=\frac{8 D \lambda T_{0}^{2}}{8 D \lambda - 9 E^{2} }
\end{equation}
and then instead of going to zero, reaches
\begin{equation}
 m^{2}_{T_1}= 2D(T_{1}^{2} - T_{0}^{2}) - \frac{9E^{2}T_{1}^{2}}{4 \lambda}
\end{equation}
 at $T=T_{1}$. For $T=T_{0}$ one has $m^{2}= 9E^{2}T_{0}^{2}/\lambda$ and for $T<T_1$
\begin{equation}
 m(T)^{2}=2D(T^{2}- T_{0}^{2}) - 6ETv + 3 \lambda v^{2},
\end{equation}
with 
\begin{equation}
v= \frac{3ET \pm \sqrt{9E^{2}T^{2} - 8D \lambda \left(T^{2} - T_{0}^{2}\right)}}{2 \lambda}.
\end{equation}
According to the now known value of the Higgs mass and to the results of non-perturbative
techniques, electroweak symmetry breaking in the Standard Model is of higher order than second \cite{Kajantie}. 
In some other extensions, such as in the Neutrino Minimal Standard Model, not only the order of the transition could be changed but also the physically viable values for the initial asymmetry between right- and left-handed particles.

From this dependence one can determine what are the temperature ranges at which chirality-flipping processes are negligible, demanding that the critical temperature obeys $\Gamma_{s}(T_{\rm out})/H(T_{\Gamma}) \approx 1$, where $H(T)$ is the Hubble parameter in the radiation dominated period given by $H\simeq1.08 \sqrt{ g_*/10.75}(T^2/M_{Pl})$ and we consider $\Gamma_s = \Gamma_H + \Gamma_{t \bar t}$ from here onwards. From here it follows that chirality-flipping processes are
out of equilibrium in the symmetric region for temperatures $T> T_{\rm out,1} \approx 2D \cdot 80$ TeV as negligible for temperatures $T_{0} < T<T_{\rm out, 2} \approx 159.5$ GeV.
Therefore, as the temperature in the symmetric region falls and approaches $T_{0} \approx 159$ GeV, chirality-flipping processes are becoming less significant.

The 1-loop approximation to high temperatures was enough to determine the Higgs thermal mass around the transition, but for the temperature and order of the transition we refer to the more detailed results obtained by lattice simulations. The discrepancy between both does not affect our analysis since they yield different results at the transition, when, as we have just seen, Higgs inverse decays are already out of equilibrium. In the remainder of this work we will use the Standard Model parameters presented above.

\subsection{Chirality flips after electroweak symmetry breaking}

\begin{figure}[tbp]
  \centering
  \includegraphics[width=0.6\textwidth]{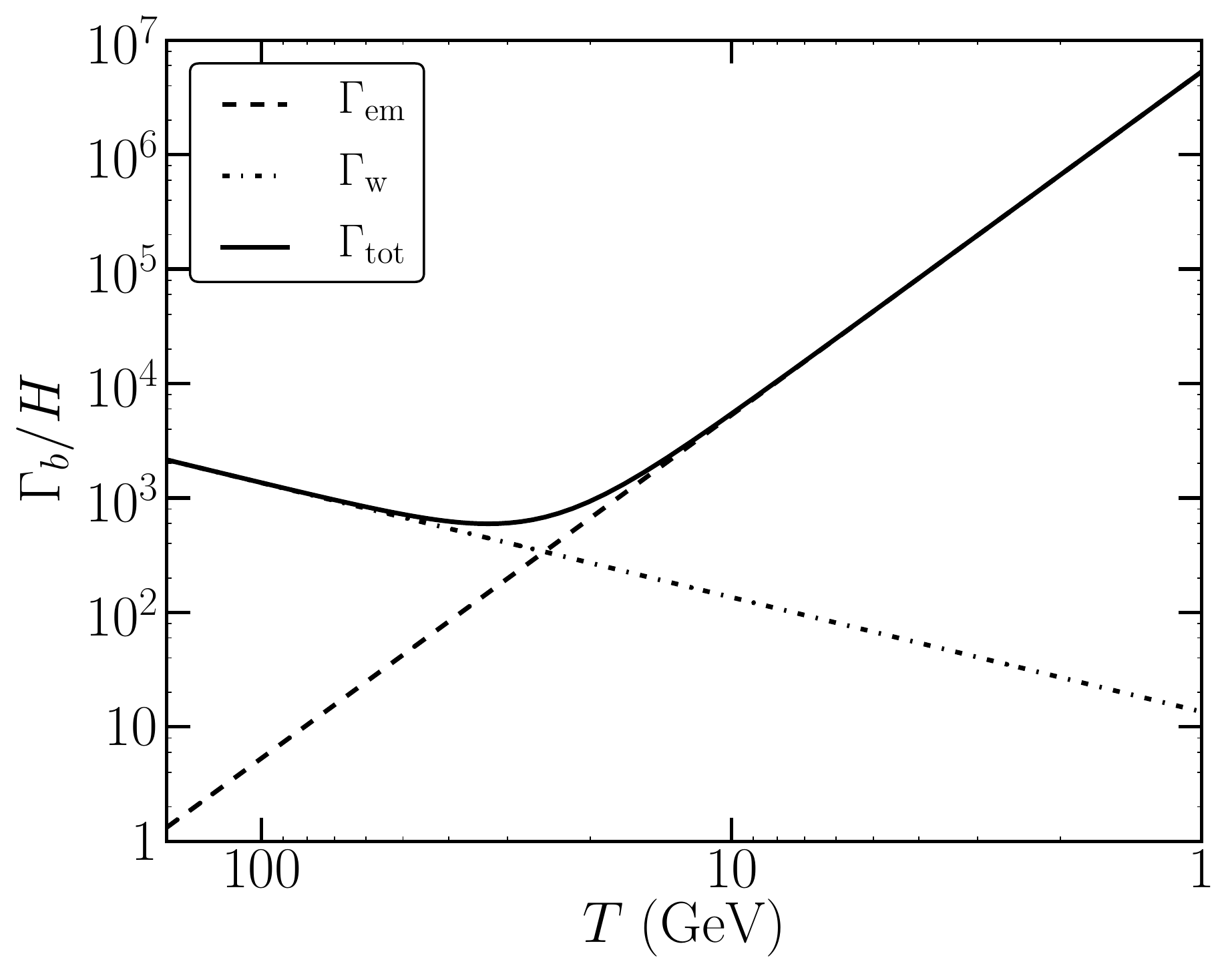}
  \caption{\label{pic:Gamma_b} Chirality flipping rate normalized to the Hubble rate in the broken phase, with $\Gamma_{\rm em}$ representing the contribution to chirality flips from electromagnetic interactions \eqref{fliping2}, $\Gamma_{\rm w}$ representing the contribution due to weak interactions \eqref{fliping1} and $\Gamma_{\rm tot} = \Gamma_{\rm em}+\Gamma_{\rm w}$ representing the total flipping rate.}
\end{figure}

Around $T_{0}$, the electroweak symmetry gets broken into the $U_{EM}(1)$ group and ordinary
electromagnetic fields take the place of the hyper magnetic fields, with the boundary condition
\begin{equation}
\textbf{B}=\textbf{B}_{Y} \cos \theta_{W}\,.
\label{boundary}
\end{equation}
The MHD equations \eqref{hmaxwell}-\eqref{hmaxwell4} and \eqref{muR} are then replaced by \eqref{Maxwell}-\eqref{c} and \eqref{mu5}, and instead of Higgs inverse decays the dominant contribution to chirality flips
now comes from weak and electromagnetic scattering processes.  The scaling of the respective cross-sections with the temperature gives the following rates
 \begin{equation}
\Gamma_{\rm w} \approx G_{F}^{2}T^4\left(\frac{m_{e}}{3T}\right)^{2}
\label{fliping1}
\end{equation}
\begin{equation}
\Gamma_{\rm em} \approx \alpha^{2} \left(\frac{m_{e}}{3T}\right)^{2}, 
\label{fliping2}
\end{equation}
where $G_{F}$ is the Fermi constant and $\alpha$ the fine-structure constant.
Both chirality flipping rates in the broken phase are shown in Fig.~\ref{pic:Gamma_b} for comparison, as well as its sum $\Gamma_{\rm tot} = \Gamma_{\rm em}+\Gamma_{\rm w}$, which is the rate we consider when mentioning $\Gamma_b$ in the following. 
From the above analysis we therefore conclude that just before the electroweak transition, 
around  $T_0 < T<T_{\rm out, 2}$, the change in asymmetry will not be significantly reduced by chirality flipping processes. On the contrary, chirality flips becomes important after the transition is completed - now dominated by
electromagnetic and weak processes. Therefore, one would expect that these features coming from the electroweak transition have significant influence on the evolution of the chiral anomaly.

\section{Analysis of the equations} \label{sec:sol_analytical}
Using the results from the previous section, and after slightly rearranging the equations, it follows that in order to investigate
MHD around the electroweak transition in the Standard Model we have to solve the following two sets
 of equations:
\begin{itemize}
 \item $T_{0}  <T< 10~{\rm TeV}$
\begin{equation}
\partial_\tau \textbf{B}^{Y}=\frac{1}{\sigma_{B}} \nabla^{2}B^{Y} - \frac{g'^{2}}{\pi^2} \frac{\mu_{R}}{\sigma_{B}} \nabla \times \textbf{B}^{Y}
\label{sym1}
\end{equation}
\begin{equation}
\frac{\dd \mu_{R}}{\dd \tau}=   \frac{g'^{2}}{8 \pi^{2}}\frac{783}{88} \frac{\dd h^{Y}}{\dd \tau} - \Gamma_s \mu_{R} 
\label{sym2}
\end{equation}
\item $T<T_{0}$ 
\begin{equation}
\partial_\tau\textbf{B}=\frac{1}{\sigma} \nabla^{2}\textbf{B} - \frac{e^{2}}{8 \pi^2} \frac{\mu_{5}}{\sigma} \nabla \times \textbf{B}
\label{b1}
\end{equation}
\begin{equation}
 \frac{\dd \mu_{5}}{\dd \tau}= 
 \frac{3e^2}{4 \pi^2}\frac{\dd h}{\dd \tau} -\Gamma_b \mu_{5}.
\label{b2}
\end{equation}
\end{itemize}
The values of the chemical potentials need obviously to be continuous while crossing 
from one region to another, with boundary condition \eqref{boundary} for fields crossing from the symmetric to the broken 
phase. We can now introduce Fourier decomposition for magnetic and hypermagnetic fields,
\begin{equation}
{\bf B}({\bf r})=\int \frac{\dd^{3}\mathbf{k}}{(2 \pi)^{3}}e^{i{\bf k} \cdot {\bf r}} {\bf B}_{\bf k},
\end{equation}
and equivalently for the modes for helicity density, ${h_{k}}$, and magnetic energy density, $\rho_{k}$. It can then be checked from the Fourier decomposition that
$\rho_{m}=\int \dd\ln k\, \rho_{k}$ and ${h}=\int \dd\ln k\,{h}_{k}$. The previous equations can suitably be rewritten in terms of helicity and magnetic
energy modes as
\begin{itemize}
 \item $T_{0}  <T< 10~{\rm TeV}$
\begin{equation} \label{eq:rhoks}
\frac{\dd\rho^{Y}_{k}}{\dd \tau}= -\frac{2k^{2}}{\sigma_{s}}\rho^{Y}_{k} - \frac{g'^{2}}{2 \pi^2} \frac{k^{2} \mu_{R}}{\sigma_{s}} h^{Y}_{k}
\end{equation}
\begin{equation}
\frac{\dd h^{Y}_{k}}{\dd \tau}= -\frac{2k^{2}}{\sigma_{s}} h^{Y}_{k} - \frac{2g'^{2}}{ \pi^2} \frac{ \mu_{R}}{\sigma_{s}} \rho^{Y}_{k}
\label{hiper-druga}
\end{equation}
\begin{equation} \label{eq:muRmodes}
\frac{\dd \mu_{R}}{\dd \tau}=   \frac{g'^{2}}{8 \pi^{2}}\frac{783}{88} \frac{\dd h^{Y}}{\dd \tau} - \Gamma_s \mu_{R} 
\end{equation}
\item $T<T_{0}$
\begin{equation} \label{eq:rhokb}
\frac{\dd \rho_{k}}{\dd \tau}=- \frac{2k^{2}}{\sigma_b} \rho_{k} - \frac{e^{2}}{16 \pi^2} \frac{k^{2} \mu_{5}}{\sigma_b}h_{k}
\end{equation}
\begin{equation} 
\frac{\dd h_{k}}{\dd \tau}=- \frac{2k^{2}}{\sigma_b} h_{k} - \frac{e^{2}}{4 \pi^2} \frac{ \mu_{5}}{\sigma_b} \rho_{k}
\label{druga}
\end{equation}
\begin{equation}
 \frac{\dd \mu_{5}}{\dd \tau}=  
 \frac{3e^2}{4 \pi^2}\frac{\dd h}{\dd \tau} -\Gamma_b \mu_{5}.
\label{mub}
\end{equation}
\end{itemize}
Apart from different chirality flipping rates in the symmetric and broken phases, an important difference comes from the fact that while
anomalous coupling is related to $\mu_{5}$ in the broken region, it is related only to $\mu_{R}$ in the symmetric region in the equations above. If we would also assume the
existence of $\mu_{L}$ in the symmetric phase and assume the presence of left-handed anomalies for the left-handed doublets (as it was done in \cite{dvornikov}) 
then the anomaly would couple to the left- and right-handed leptons separately instead of to their difference, $\mu_{5}$, as in the symmetric phase. This means that it is in principle
possible to start even from $\mu_{5}=0$ in the symmetric phase and then obtain its growth through this type of coupling before the
electroweak transition, after which the produced $\mu_{5}$ would couple to the ordinary magnetic field. Yet, introducing $\mu_{L}$
would actually violate the equilibrium of five chemical potentials for five conserved charges, and one would also need to take 
into account sphaleron processes, which couple only to left-handed particles, and violate lepton and baryon number. In the present work we will follow the approach which assumes that in the symmetric phase only $\mu_{R}$ is non-vanishing, as put forward in \cite{Giovannini3}.
This is in accord with scenarios where the baryon asymmetry of the universe comes from leptogenesis and is stored in right-handed 
electrons before the electroweak transition, in which there is no asymmetry between left-handed particles, i.e. $\mu_{L}=0$ \cite{Davidson}.

Another difference between the symmetric and broken phases are the values of the conductivities $\sigma_{s}$ and $\sigma_{b}$, respectively, since before the electroweak
breaking interactions between leptons, $W^{\pm}$ and $Z^{0}$ reduce the conductivity. However, it seems that this difference
is not significant and can be approximated by $\sigma_{s} \approx \sigma_{b} \cdot \cos^{4} \theta_W$ \cite{Baym}.

Since equations in different phases have the same form mathematically, only with different coefficients and flipping rates, in order to analyse
them together we will introduce the notation 
\begin{equation}
\begin{split}
c_{1}= \frac{g'^{2}}{2 \pi^2} \frac{1}{\sigma_{s}}, & \; \; c_{2}=\frac{e^{2}}{16 \pi^2} \frac{1}{\sigma_{b}}, \\  
c_{3}= \frac{g'^{2}}{8 \pi^{2}}\frac{783}{88}, & \; \; c_{4}=\frac{3 e^2}{4 \pi^2}.
\end{split}
\end{equation}
Although the presented modified magnetohydrodynamical equations are not solvable analytically we can can still obtain some interesting conclusions by analysing them in certain regimes.
It can be seen that the evolution of the chiral anomaly will in general depend on the relative strength of the (hyper)helicity time change and chirality flips. 
We can estimate the value of the initial magnetic energy density necessary to prevent the fast damping of the asymmetry before the transition by requiring $\dd\mu_5/\dd \tau\simeq 0$, which leads to $\mu_5\cdot \Gamma_s/(c_3|\dd h/\dd \tau|)\simeq 1$.  Since we are at the moment interested in the limiting regimes of the evolution of $\mu_{5}$, we can for simplicity take the maximal helical case, $h_{k}=2 \rho_{k}/k$, and approximate the spectral distribution as $h_{k}(\tau)=h(\tau)(k/k_{max})^n$, for $k\leq k_{max}$, with $k_{max}$ corresponding to the shortest length scale. This distribution is, strictly speaking, valid only for the initial moment, but for the considered regimes it will not be significantly modified and can be taken in the analytical treatment as an approximation.
By virtue of integrating \eqref{hiper-druga}, we obtain
\begin{equation} \label{eq:h_dampthr}
h \simeq \frac{\mu_5 \Gamma_s}{c_3}\frac{|k_{max}^{n+1}-k_{min}^{n+1}|}{(n+1)\left[-\frac{2}{\sigma_s(n+3)}(k_{max}^{n+3}-k_{min}^{n+3})-\frac{g'^2 \mu_5}{\pi^2\sigma_s(n+2)}(k_{max}^{n+2}-k_{min}^{n+2})\right]}
\end{equation}
and integrating the initial distribution $\rho_k$, yields 
\begin{equation} \label{eq:rho_dampthr}
 \rho_m \simeq \frac{(n+1)}{2(n+2)}\frac{k_{max}^{n+2}-k_{min}^{n+2}}{k_{max}^{n+1}-k_{min}^{n+1}}h.
\end{equation}

For sufficiently small (hyper)field energy densities and helicities, one can take $(\dd h/ \dd \tau) \approx 0$ and the chiral anomaly evolution
is then given by an exponential decay
\begin{equation} \label{eq:mu5Rapprox}
\mu_{R,5}\approx \mu^{0}_{R,5} \exp\left(-\int \Gamma_{s,b}(\tau) \dd\tau\right), 
\end{equation}
where $\mu_{R,5}$ denotes the chemical potential of the chiral asymmetry in the electroweak phase and in the broken phase, respectively. On the other hand, for sufficiently strong fields we have
$\mu_{R,5} \cdot \Gamma_{s,b}/(c_{3,4} |\dd h/\dd\tau|) \ll 1$. Then we obtain $\mu_{R,5} \approx c_{3,4} h +c$, with $c$ a constant set by the initial conditions. Solving the remaining differential equation under the aforementioned assumptions we get
\begin{equation}
\mu_{R,5} \approx \frac{1}{2b}\left[d \cdot \tanh \left(\frac{d \tau}{2}\right) -f\right],
\end{equation}
\begin{equation}
\rho_{m} \approx\frac{1}{2 c_{1,2}} \frac{b}{ c_{3,4}}[\mu_{R,5} - c],
\end{equation}
with the coefficients
\begin{equation}
\begin{split}
b=\frac{n+1}{n+2} c_{1,2} \frac{k_{max}^{n+2} -k_{min}^{n+2} }{k_{max}^{n+1} -k_{min}^{n+1}}, &\; \;\; e=\frac{2(n+1)}{(n+3) \sigma_{s,b}}  \frac{k_{max}^{n+3} -k_{min}^{n+3} }{k_{max}^{n+1} -k_{min}^{n+1}}, \\
d=\sqrt{(b \cdot c - e)^{2} + 4b\cdot  c \cdot e}, &\; \;\; f=b \cdot c - e ,
\end{split}
\end{equation}
with $k_{min}$ denoting the largest length scale. 

In this regime $\mu_{R,5}$ will therefore grow due to its coupling to the (hyper)magnetic field. We expect that
this regime correctly describes the evolution of $\mu_{5}$ near the electroweak transition, as well as for temperatures higher than $T_{\rm out,1}$, under the assumption that hypermagnetic fields existed at that period. On the other hand, for temperatures lower than the electroweak transition, spin flipping processes become dominant and one expects that $\mu_{5}$ will be eventually exponentially damped. At the electroweak transition we cross from a regime of negligible chirality flipping rates to one of relatively high flipping rates -- therefore we expect a sharp jump in $\mu_{5}$ due to this change of regime, with $\mu_{5}$ finally approaching the limit 
\eqref{eq:mu5Rapprox} at lower temperatures. Due to the behavior of the spin flip rate in the broken phase (see Fig.~\ref{pic:Gamma_b}), and assuming the existence of magnetic fields, this regime is not reached instantly after the transition, but $\mu_{5}$ can actually increase for some time even in the broken region. 
This happens until it reaches a maximum that occurs around the corresponding minimum of chirality flipping rate, which can be calculated from (\ref{fliping1}) and (\ref{fliping2}) as $T(\mu_{5}^{max})\approx(\alpha/G_{F})^{1/2} \sim$~GeV.
We will indeed see in the next section that the general numerical solution of the system confirms our analytical predictions.

A third important and simple limit is that of a stationary chiral anomaly potential, $\mu_{R,5}(t)=const.\equiv \mu_{R,5}^{st}$. In order to
achieve this solution, the change in helicity needs to compensate the loss from spin flips according to $\delta h = \mu_{R,5}^{st}/c_{3,4} 
\int\Gamma_{s,b}(\tau) \dd\tau$. Then, taking again for simplicity the maximal helical case, one can obtain the
exact solution for the helicity modes
\begin{equation}
h_{k}=h_{k}^{0} \exp\left[-(\tau - \tau_{0})\cdot\left(\frac{2k^{2}}{\sigma_{s,b}}+c_{1,2}k \mu_{R,5}^{st}\right)\right].   
\end{equation}
If $\mu_{R,5}^{st}$ and $h_{k}$ have opposite signs then a stationary solution, which corresponds to maximal helicity (and also to magnetic energy in the maximal helical case),
is reached for the mode 
\begin{equation} \label{eq:k5}
 k_{R,5}=  \frac{c_{1,2} \sigma_{s,b}}{2} \lvert \mu_{R,5}\rvert.
\end{equation}
 All modes for which $k<k_{R,5}$ will then be growing, and
in the opposite case decaying. This corresponds to the transfer of helicity (and energy) from smaller to larger length scales -- i.e. an inverse cascade.
We see that in the presence of the chiral anomaly, the effect of an inverse cascade can therefore be obtained even without a velocity field. When $\mu_{5}$ is not
stationary, the presented analytical solution is no longer valid because it becomes time dependent, but the qualitative analysis remains. Now $k_{R,5}$ is a function of time, and therefore the character of a given mode (growing, decaying or stationary)
will generally change with time accordingly. Being related to the chiral asymmetry $\mu_{R,5}$ by \eqref{eq:k5},
we expect that the helicity maximum will shift towards smaller length scales around the electroweak transition, since $\mu_{R}$ is 
increasing in that region, and that it will, on the other hand, shift towards larger length scales at temperatures lower than the electroweak transition.
 
It is interesting and important to note that modified MHD equations lead to the existence of a non-vanishing helicity of the magnetic field, even if 
no initial helicity was present. This can be easily seen in the following way: assuming no helicity, energy modes at first
just decay, $\rho_{k}=\rho_{0} \exp(-2k^{2} \tau/\sigma_{s,b})$. But on the other hand, the time change of the helicity is still non-zero due to the presence of $\mu_{5}$, and this leads to the growth of helicity density under the assumption that $\mu_{5}$ has an opposite sign.
One then gets that the helicity density produced in the fast growth regime in the symmetric phase is given by 
\begin{equation}
\lvert h_{pr} \rvert=  \frac{2g'^{2}}{ \pi^{2} \sigma_{s}}\left|  \int \dd\tau \int \dd k\, k^{2}\mu_{R}(\tau) \rho_{k}(0)e^{-\frac{2k^{2} \tau}{\sigma_{s}}}  \right|.
\end{equation}
The helicity in the right-hand side of \eqref{hiper-druga} and \eqref{druga} will grow in approximately the same fashion as the term containing it stays much smaller than the term with the energy density. When this is not the case, one needs to solve the full set of coupled equations for energy and helicity densities. Starting from a vanishing helicity we therefore expect to
have its fast growth until the term containing it reaches a level comparable to the magnetic energy term, with a further less dramatical evolution closely related to the evolution of magnetic energy. In the next section we will also confirm this qualitative conclusion with numerical simulations.   

One of the important consequences of the presence of helicity in cosmological magnetic fields is that it can lead to an increase
of the correlation length. An important problem with the scenarios of primordial magnetic field generation is the small correlation length
of the fields which, if not produced during inflation, needs to be smaller than the Hubble radius at the period of their creation. This is in sharp contrast with the correlation lengths
of observed fields, which can be even of the order of Mpc \cite{Giovannini2}. However, magnetic helicity in the presence of turbulence is known to lead to the development of magnetic structures at progressively larger sizes as a consequence of the inverse cascade that occurs. In this 
framework, the correlation length can grow considerably \cite{brandenburg, camp}. The evolution of the chiral anomaly around the electroweak
transition, naturally leading to the creation of helicity, could therefore also influence the growth of the magnetic field correlation 
length if turbulence develops. Also, the presence of helicity could change constraints on the amplitude of primordial magnetic fields
from gravitational wave production \cite{caprini}. 

Conversely, if one assumes an initial helicity and no initial asymmetry between left- and right- handed leptons, by a similar logic it follows
that helicity will at first just decay, which will lead to a non-vanishing time derivative of $\mu_{5}$ -- therefore producing a
chiral asymmetry associated with the chemical potential
\begin{equation} \label{eq:mu5pr}
\mu_{5}^{pr} =- \frac{c_{3,4}}{\sigma_{s,b}}  \int \dd\tau \int \dd k\, 2k^{2} h_{k}(\tau). 
\end{equation}
This creation of chiral asymmetry due to an helical magnetic field could be important to understand the evolution of particle distribution in
the early Universe, due to the different couplings between left- and right-handed particles, and can therefore be relevant for different 
baryogenesis models. This is of special interest for some
recent models where it is suggested that helical hypermagnetic fields
could produce the baryon asymmetry of the universe~\cite{Kohei}.

\section{Numerical solutions}  \label{sec:sol_numerical}

After discussing the analytical limits of the modified MHD equations for the evolution of the magnetic energy, helicity and chiral chemical potential, we now solve the system of differential equations \eqref{eq:rhoks}-\eqref{mub} numerically, both before and after the electroweak transition, with the aim of understanding the implications of the presence of an electron chiral chemical potential $\mu_{R,5}$ around this transition. We use non-conformal units in this section for the convenience of the reader. 

As motivated in \S\ref{sec:sol_analytical}, we take the inital chemical potential of left-handed electrons before electroweak symmetry breaking to be zero, $\mu_L/T =0$,  and as a rough estimate of the chiral asymmetry present at that epoch, we consider the asymmetry between baryons and antibaryons, setting $\mu_5^0/T= 10^{-9}$ at the initial temperature $T=300$ GeV, chosen for conviniently showing the behaviors before the transition.  We solve our system of equations for a spectrum approximated to a number of $i=1,10$ modes $k_i =k_{min} 2^{i-1}$, with the chosen $k_{min}/T=10^{-10}$. We note that since $H \ll k_{ min}$, we are always dealing with length scales smaller than the Hubble length. 
For the initial magnetic energy, we assume that seed magnetic fields in the early universe are relatively small compared to the total initial energy density $\rho_{\rm tot}= \pi^2 g_*T^4/30$ and we denote the ratio between the magnetic energy density and total energy density $\Omega_{\rm mag} = \rho_m/\rho_{\rm tot}$. We assume an initial power spectrum of the type $\rho_k^{0}/\rho_{\rm tot}=\Omega_{\rm mag}^0 5k^5/(k_{max}^5-k_{min}^5)$, with  $k_{min} < k < k_{max}$,
for the magnetic energy density.
The magnetic helicity is the source of the link between the chiral anomaly and MHD, as we know from \eqref{mu5} and \eqref{muR}, and we choose two limiting cases -- having an initially vanishing and maximal helicity density, $h_{max}(k)=2 \rho_k^{(Y)}/k$. We consider either $h_0=0$ (represented in the Figures in green) or $h_0=h_{max}^0(k)$ (represented in the Figures in red).

\subsection{Evolution of chiral chemical potential}  \label{sec:mu5}

\begin{figure}[tbp]
  \centering
 \includegraphics[width=.51\textwidth]{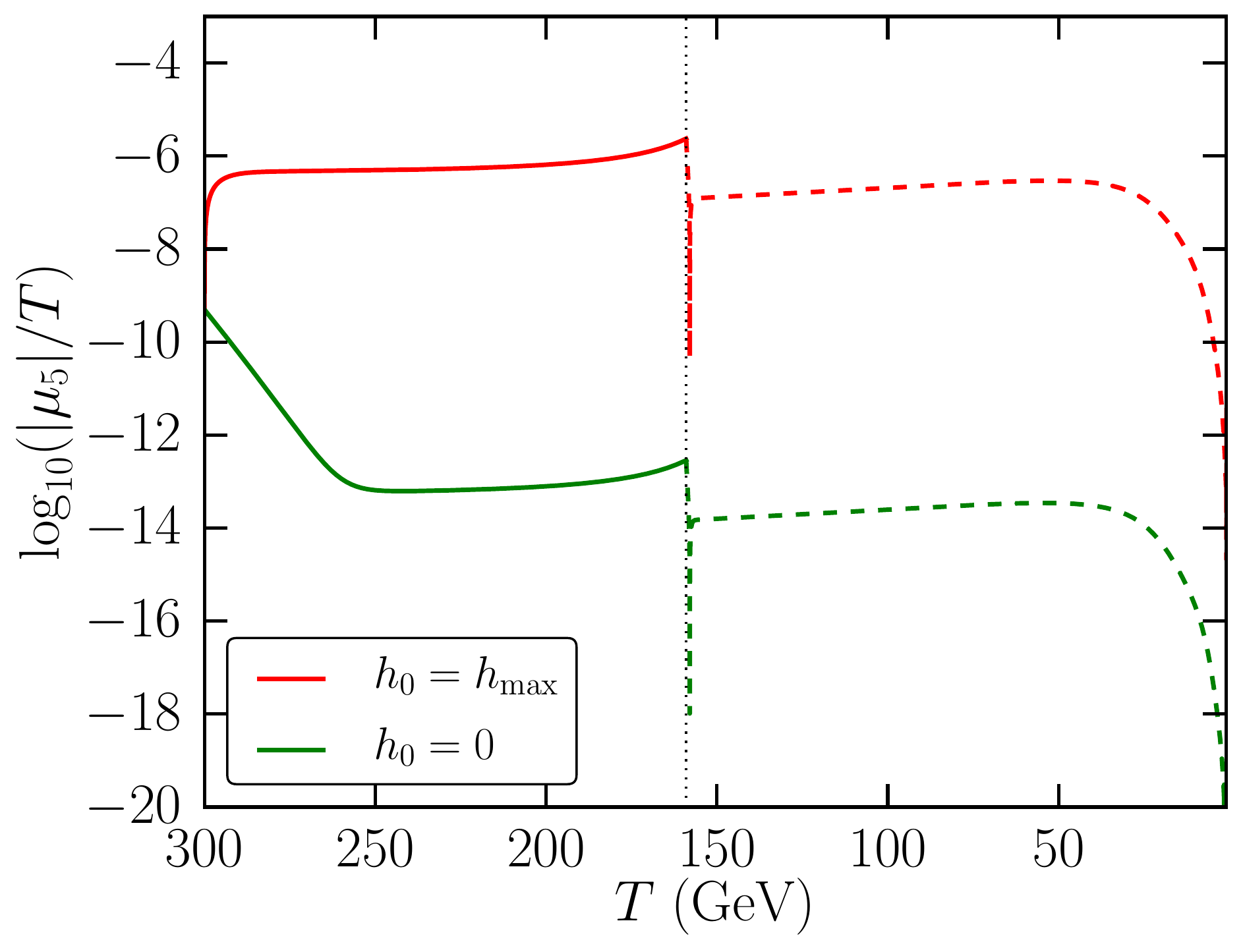}
 \hfill
 \includegraphics[width=.47\textwidth]{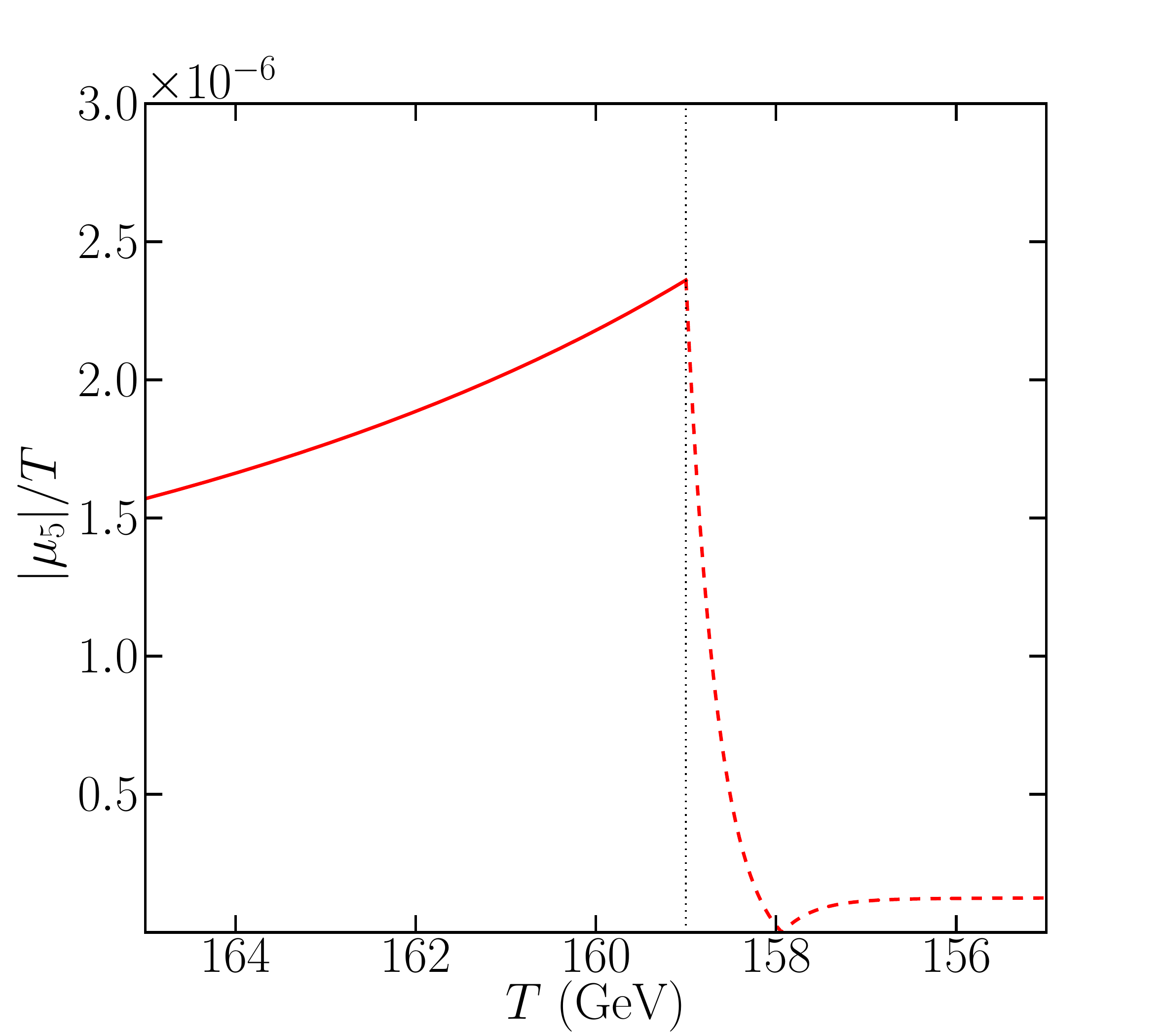}
 \caption{ \label{pic:mu5} Evolution of the logarithm of the chiral chemical potential $\log_{10}(|\mu_{5}|/T)$ with temperature, before (solid lines) and after (dashed lines) the electroweak transition, with $\Omega_{\rm mag}^0=10^{-10}$ and $\mu_5^0/T=10^{-9}$, beginning at $T=$300 GeV, for the minimal initial helicity density $h^Y_0=0$ (in green) and maximal $h^Y_0=h_{\rm max}$ (in red), on the left-hand side. Zoom around the transition for the initially maximal helical case on the right-hand side.}
\end{figure}

In Figure~\ref{pic:mu5} the behavior of the electron chiral chemical potential is shown, according to \eqref{eq:muRmodes} and \eqref{mub}, both for the initially non-helical and maximal helicity cases. One can see that the electroweak transition in the Standard model can in principle have a large influence on the evolution of $\mu_{5}$ leading to its very rapid decrease around the transition temperature. This demonstrates the possible influence of the electroweak transition on the evolution of the chiral asymmetry -- and subsequently on the magnetohydrodynammics -- of the early universe. 
The general features of the evolution of $\mu_{5}$ are in accordance with the conclusions of the analytical discussion in the last section, namely that starting from a temperature above the electroweak transition, here chosen to be $T=300$ GeV and assuming $\mu_{L}=0$, $|\mu_{5}|=|-\mu_R/2|$ grows while the temperature approaches $T_{0}\simeq 159$ GeV due to the 
decrease of the chirality flipping rate $\Gamma_s$ (see \eqref{eq:Gamma_s}, \eqref{eq:ttbar}
and Fig.~\ref{pic:Gamma_s}). The value of the initial temperature is not having a significant influence on the evolution of $\mu_{5}$ and the initial temperature $T=300$ GeV. The right-hand side of Figure~\ref{pic:mu5} zooms into the phase transition to better illustrate the change of behavior that $|\mu_{5}|$ undergoes in this region. At the transition crossing (represented by the vertical line), $|\mu_{5}|$ suddenly falls due to the activation of weak interaction and electromagnetic spin flipping processes, determined now by the non-vanishing electron mass, according to \eqref{fliping1} and \eqref{fliping2}, respectively. After the transition, $|\mu_{5}|$ continues to grow slowly in the broken phase until it reaches a maximum, determined by the mimimum of flipping rates which occurs around 40 GeV. After that, $|\mu_{5}|$ rapidly decays and eventually reaches the limit of exponential decay. This regime was studied in \cite{Boyarsky}, but where only electromagnetic processes were taken into account in the chirality flipping rate.

The difference between red and green lines shows a range of about 6 orders of magnitude of possible values that $|\mu_{5}|$ can take depending on the initial value chosen for the helicity density, which shows the impact that helicity has on the chiral asymmetry magnitude.

\begin{figure}[tbp]
  \centering
 \includegraphics[width=.51\textwidth]{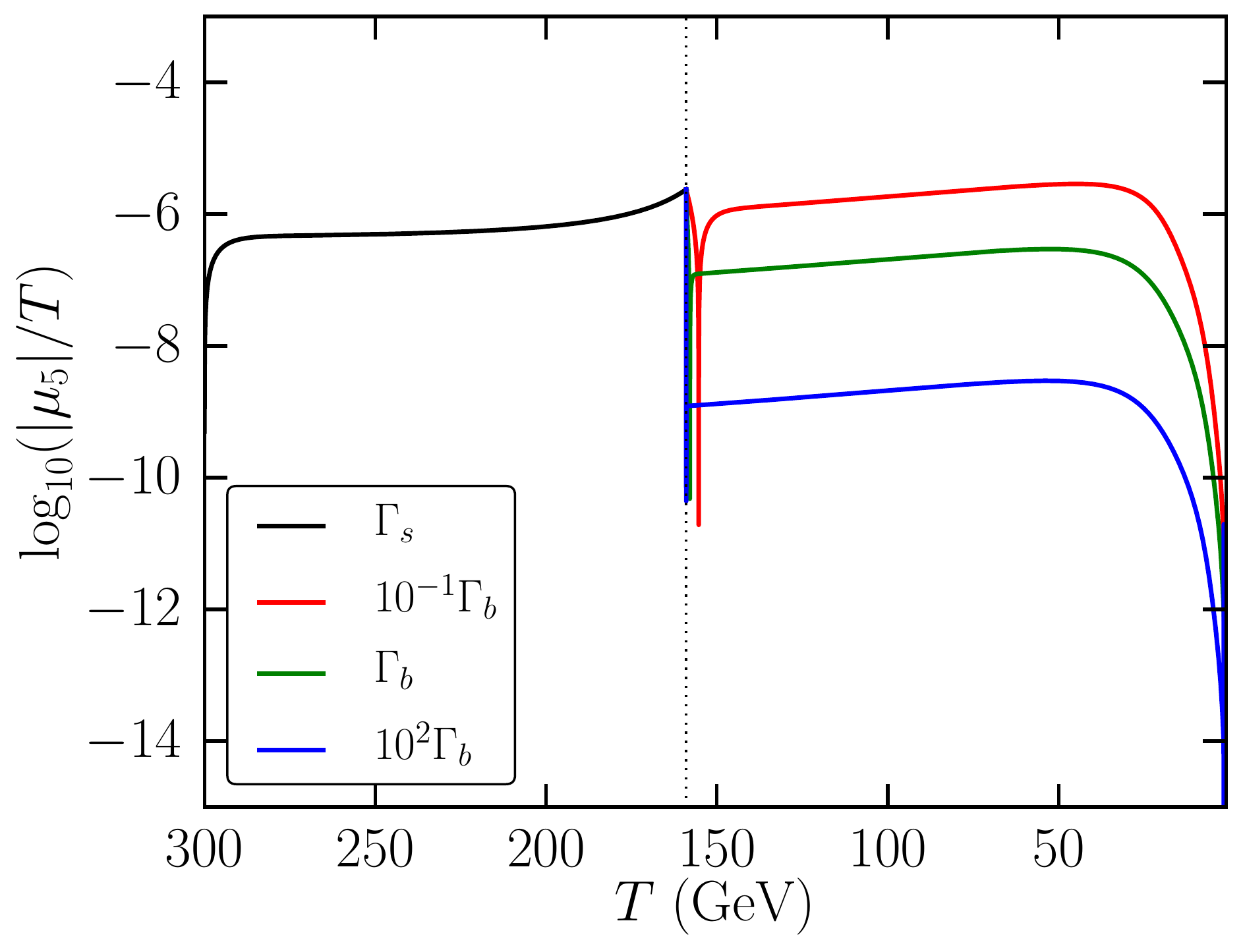}
 \caption{ \label{pic:mu5_gamma} Evolution of the logarithm of the chiral chemical potential $\log_{10}(|\mu_{5}|/T)$ with temperature, with $\Omega_{\rm mag}^0=10^{-10}$ and $\mu_5^0/T=10^{-9}$, for the maximal initial helicity density $h^Y_0=h_{\rm max}$, for different modified values of chirality flipping rates in the broken phase.}
\end{figure}

Shortly after the transition, since gauge bosons are still light, one can question how justified the estimates of electromagnetic and weak chirality flipping rates are. If we still take into account the Yukawa interactions, their small contribution in the beginning of the broken phase is negligible, not changing the total magnitude of the flipping rates. Moreover our analysis does not qualitatively dependent on the particular values of the flipping rate just shortly after the transition: in Fig.~\ref{pic:mu5_gamma} we show this independence modelling the behaviour of the evolution of $\mu_5$ by modifying the chirality flipping rates over several orders of magnitude. Therefore, the considered flipping rates are a reasonably good approximation and in case they are higher due to additional contributions, their effect in the asymmetry evolution will just be more significant.

\subsection{Evolution of the magnetic energy} \label{sec:rho_m}

\begin{figure}[tbp]
  \centering
 \includegraphics[width=0.6\textwidth]{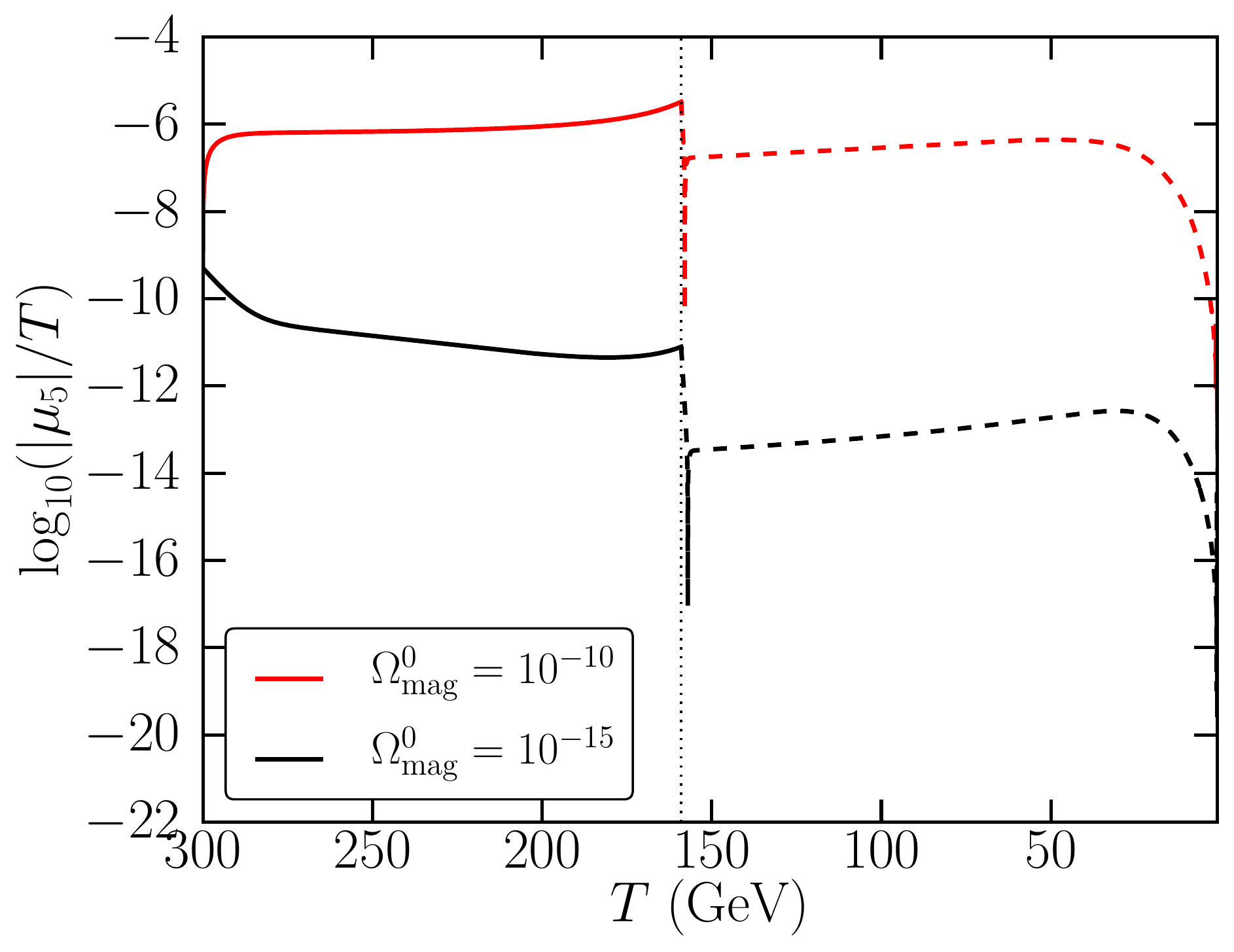}
 \caption{\label{pic:2rB} Evolution of the logarithm of the chiral chemical potential $\log_{10}(|\mu_{5}|/T)$ with temperature, with $\Omega_{\rm mag}^0=10^{-10}$ in red, as in Fig.~\ref{pic:mu5}, and $\Omega_{\rm mag}^0=10^{-15}$ in black, for a maximal initial helicity density and with $\mu_5^0/T=10^{-9}$. }
\end{figure}

The initial ratio between the magnetic and the total radiation energy densities, $\Omega_{\rm mag}^0$, represents an important quantity since it determines the regimes of evolution of $\mu_5$ and $\Omega_{\rm mag}$, together with the chirality flipping rate. Therefore it also determines the temperature at which $\mu_5$ gets damped to negligible values in the broken phase. The value of $\Omega_{\rm mag}^0$ has to be significantly smaller than unity, implying that the magnetic energy is much smaller than the radiation energy, such that it can be neglected in terms of the expansion of the universe. On the other hand, magnetic fields should already be strong enough such that the exponential decay of $\mu_{5}$ before the transition is prevented. 
This is shown in Figure~\ref{pic:2rB}, which represents the chiral asymmetry evolution for two different values of the initial magnetic energy density. This figure allows us to compare the growth of $\mu_5$ in the symmetric region when $\Omega_{\rm mag}^0=10^{-10}$ with the decay from its initial value when $\Omega_{\rm mag}^0=10^{-15}$. Taking the values of $\mu_5$ and $\Gamma_s$ at 300 GeV, through \eqref{eq:h_dampthr} we can estimate the minimal value required for the initial magnetic energy in order for the instability to survive until the transition without being significantly damped. Inserting these values in \eqref{eq:rho_dampthr} we find that for $\Omega_{\rm mag}^0\gtrsim 10^{-15}$, $\mu_5^0$ will not suffer decay, which is in complete agreement with the presented numerical result.

\begin{figure}[tbp]
  \centering
 \includegraphics[width=0.6\textwidth]{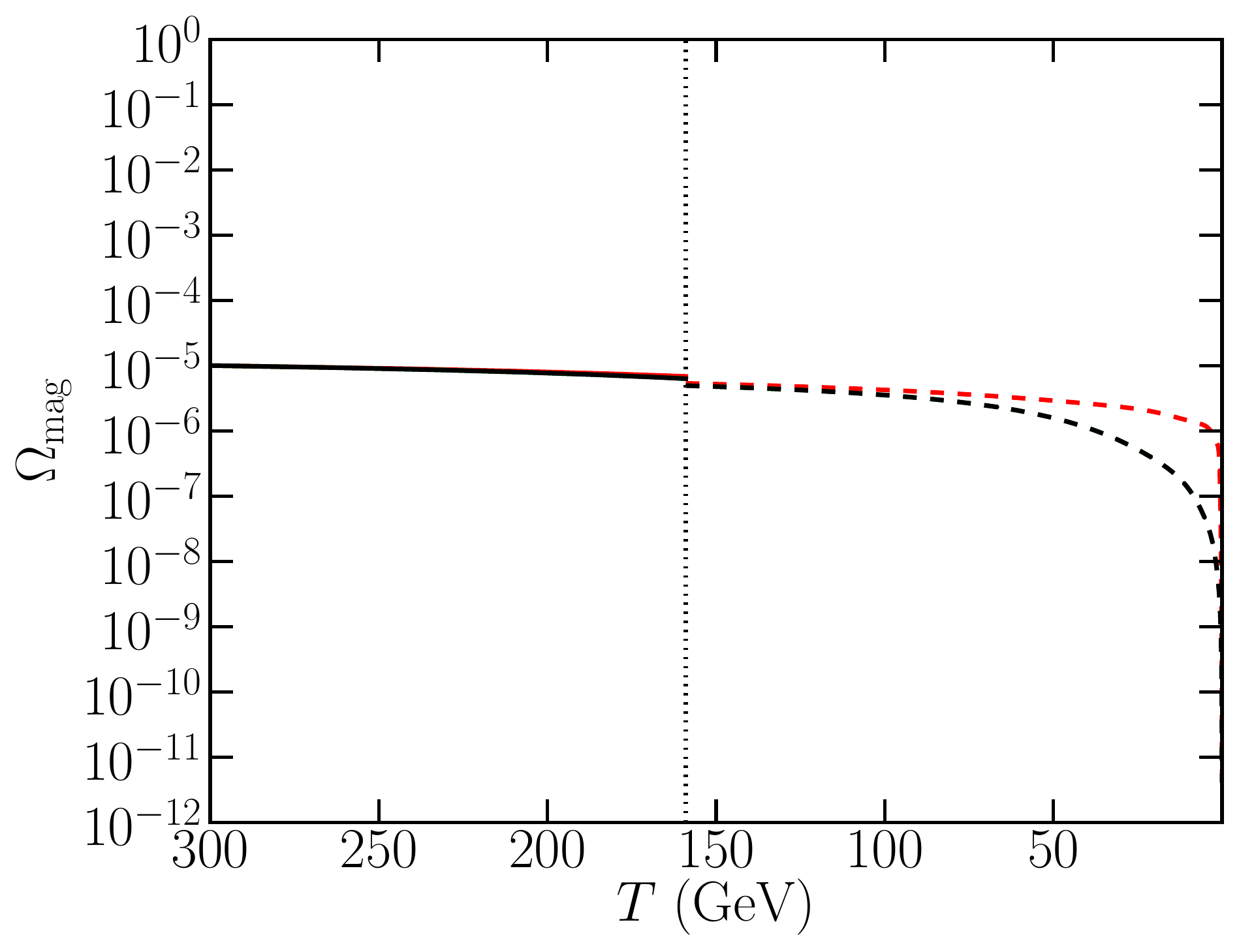}
 \caption{\label{pic:energy}  Evolution of the magnetic energy density normalized to the total energy density, $\Omega_{\rm mag}^Y$ before (solid lines) and $\Omega_{\rm mag}$ after (dashed lines) the electroweak phase transition with respect to temperature for $\Omega_{\rm mag}^0=10^{-5}$ and $\mu_5^0/T=10^{-9}$, in red. The curves in black represent the evolution of the magnetic energy density in the absence of $\mu_5$. }
\end{figure}

Figure~\ref{pic:energy} shows in red the magnetic energy evolution in the region preceeding the electroweak phase transition, following \eqref{eq:rhoks}, and following \eqref{eq:rhokb} for $T<T_0$. The curve in black shows the magnetic energy evolution in case the chiral asymmetry was not present, i.e. if the magnetic fields would only be subject to resistive damping.  
One observes that there is no significant growth of magnetic energy in the analysed regions, not even in the short interval around the phase transition where
flipping processes are negligible. This is not surprising since even in the regime of no
chirality flips we have $\delta h= \delta \mu_{R}/c_{3}$, assuming the maximally helical case. For $\delta (|\mu_{5}|/T) \sim 10^{-6}$ around the transition, obtained from solving the evolution equation, whose solution is shown in Fig.~\ref{pic:mu5}, this leads to an insignificant
growth in helicity/magnetic energy. 
Therefore a possible transfer of chiral energy into magnetic energy would bring no significant field enhancement. This can also be understood by comparing the typical magnetic field growth time scale and the electroweak transition time scale. Taking again the maximal helical case, if the conditions for the growth of a mode are satisfied ($k<k_{5}$, opposite sign between $\mu_{5}$ and $h^{(Y)}$), we see from (\ref{hiper-druga}) that growth will be described by the term $ \Gamma_g \equiv g'^{2} k \mu_{R} /(\pi^{2} \sigma_{s})$. We expect the characteristic growth time
to be $\tau_{g} = 1/\Gamma_{g}$, where we can take $k \approx k_{5}$ since, as we have shown, most of the energy would be stored around this mode. Then for $\mu_{5}/T \sim 10^{-6}$, and introducing the transition time scale, $\tau_{\rm tr}=1/\delta T$, $\delta T=0.5$ GeV, one obtains $\tau_{\rm tr}/\tau_{g} \simeq 10^{-3}$. This shows that the time scale of the transition is too short for significant growth of field strength. Even if one would assume much higher -- and therefore less physically natural -- initial values for $\mu_{5}$, but which could occur in extensions of the Standard model, one would still need to also assume the existence of strong seed magnetic fields to support the anomaly, which would make the logic of significant field amplification in this context circular. 
We therefore conclude that it does not seem likely that any significant magnetic field growth occurred at the electroweak transition,
which is a cross over or a second order transition, even if we take into account modifications of the MHD equations due to the chiral magnetic effect. 

Additionally, Figure~\ref{pic:energy} shows the difference between ordinary resistive damping and in the presence of $\mu_5$. The decay obtained in the studied modified MHD is thus slower due to the anomaly effect and as the asymmetry gets eventually damped, it can be seen that after the transition the evolution approaches the limit of an exponential decay.

\subsection{Evolution of helicity density} \label{sec:hel}

\begin{figure}[tbp]
  \centering
 \includegraphics[width=.55\textwidth]{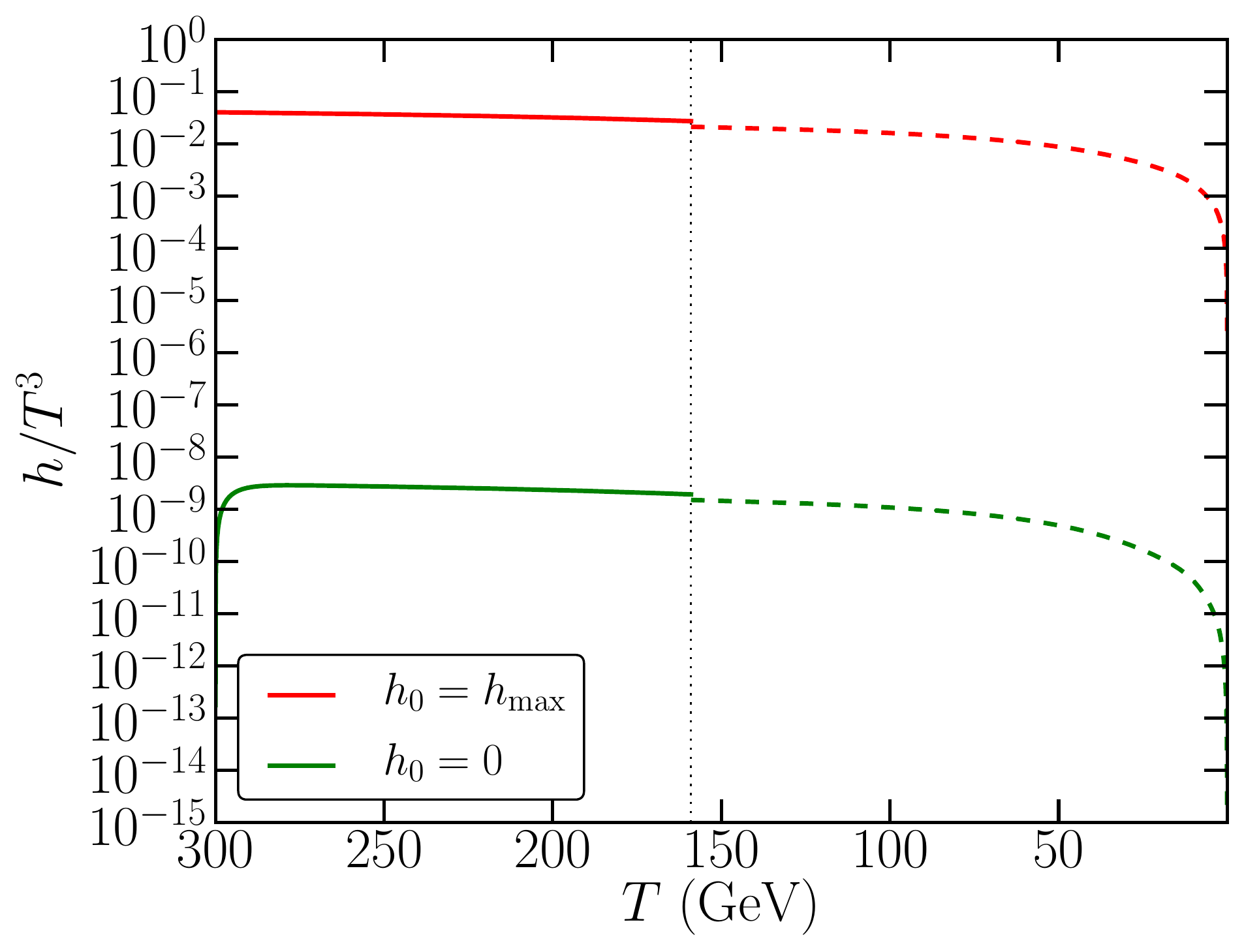}
 \hfill
 \includegraphics[width=.43\textwidth]{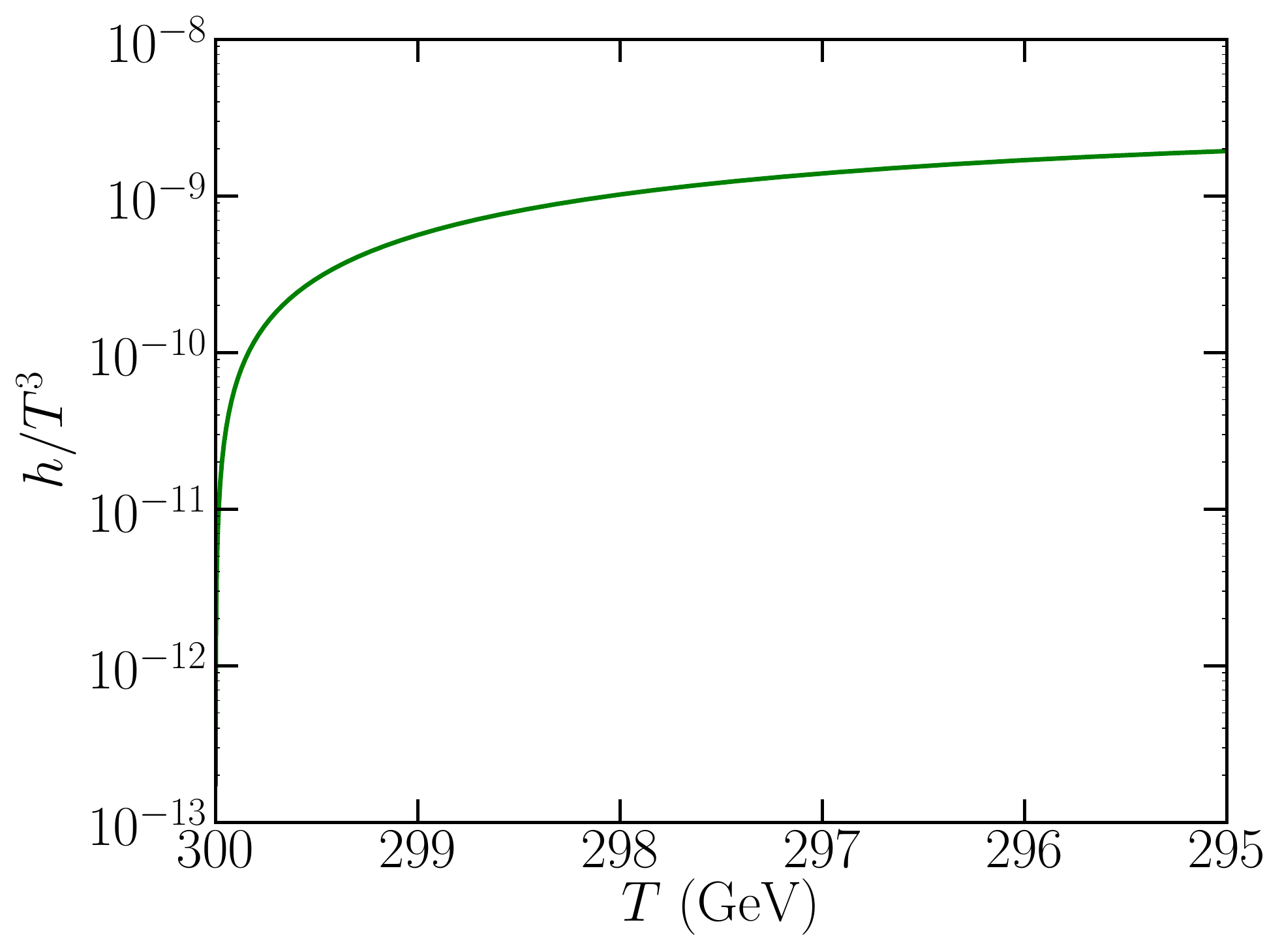}
 \caption{ \label{pic:helicity} Evolution of helicity density, $h^{Y}/T^3$ before (solid lines) the electroweak transition and $h/T^3$ after (dashed lines) the transition for the minimal initial helicity density $h^Y_0=0$ and maximal $h^Y_0=h_{\rm max}$, with respect to temperature and using $\Omega_{\rm mag}^0=10^{-10}$ and $\mu_5^0/T=10^{-9}$ on the left-hand side. Zoom on the high temperature region to show the helicity growth from $h^Y_0=0$ to its later stable value on the right-hand side.}
\end{figure}

Figure~\ref{pic:helicity} shows the helicity evolution according to \eqref{hiper-druga} before the phase transition and dictated by \eqref{druga} afterwards. Despite the absence of significant magnetic energy amplification, the chiral anomaly leads to the creation of non-vanishing helicity even if no initial helicity was present, which is in accordance with the analytical discussion in the previous section and visible in particular on the right-hand side of Fig.~\ref{pic:helicity}. It is therefore not a necessary condition to assume that magnetic fields are initially helical in order for a chiral asymmetry to have influenced the electroweak transition and the magnetohydrodynamics of the early universe. The difference between both cases is that an initially non-helical field will not evolve into a fully helical one, implying that the values of helicity density and $\mu_{R,5}$ will in this case be lower. The reason for this is the fact that a lower helicity means that spin flipping processes will have a stronger influence.

\section{Summary and conclusions} \label{sec:conclusions}

In the early universe, phase transitions offer a privileged environment to investigate possible cosmological mechanisms that are important for the understanding of the evolution of primordial magnetic fields. Motivated by recent proposals that the chiral anomaly may play a role in cosmological magnetic fields, we have analysed this effect specifically at the electroweak transition, assuming no extensions of the Standard Model. Some possible extensions, which could make the electroweak transition a first order one -- 
and therefore lead to MHD turbulence via bubble collisions -- include modifications of the Higgs potential with terms of higher order, the two-Higgs doublet model, the minimal supersymmetric standard model and next to-minimal supersymmetric standard model. In the context of a higher order transition it is also interesting to ask whether it is possible to obtain significant field amplification, which could contribute to the generation of primordial magnetic fields.
By writing down the equations governing the evolution of hyper- and ordinary magnetic fields in a plasma with a chiral imbalance ($n_R\neq n_L$), as well as including the processes that can change the chiral number, we were able to study the behavior of helicity, magnetic energy and chiral chemical potential. Unlike a first order phase transition, higher order transitions do not lead to any direct cosmological consequences,  such as magnetic field amplification or gravitational wave production.
However, we have shown that the Standard Model electroweak transition can still have consequences for the magnetohydrodynamics of the universe when the chiral anomaly is taken into account.

In the typical conditions of the early universe and for the initial magnetic energy considered, a chiral chemical potential $\mu_5$ present before the electroweak transition will survive the crossover transition and further down to temperatures of the order of tens of GeV, when the chirality flipping rate dominates its evolution and $\mu_5$ decays exponentially. The different nature of the interactions before and after the transition give rise to different chirality flipping rates that greatly change the behavior of the chiral chemical potential during the transition. Assuming an initial asymmetry of $\mu_5^0/T=10^{-9}$ at 300 GeV, for maximal helicity the typical asymmetry values lie around $10^{-6}T$ in the vicinity of the transition and for an initially vanishing helicity, around $10^{-14}T$.

Regarding the evolution of the magnetic field, we have shown that the magnetic fields will not be significantly affected by the presence of a $\mu_{R,5}$ under physically reasonable initial conditions, namely that they will not undergo any notable amplification but their decay will be slowed down.  On the other hand the addition of $\mu_{R,5}$ to the MHD equations naturally leads to the creation of helical magnetic fields, even if they were initially non-helical. 

The study of these modified MHD equations is therefore relevant for a number of reasons. First, it provides a more accurate treatment of the early universe plasma by extending standard MHD to take into consideration the chiral nature of the particles composing it. It also allows us to disregard the standard model electroweak transition as a source of cosmological magnetic field enhancement via the chiral anomaly. At the same time this shows that, although no growth may take place, non-helical magnetic fields are transformed into helical magnetic fields due to the chiral anomaly, which can be of interest since the presence of helicity in the later evolution of cosmological magnetic fields can impact the growth of the magnetic field correlation length.
On the other hand, if we assume that cosmological magnetic fields were originally helical, then modified MHD equations 
lead to the creation of an asymmetry between left- and right-handed particles, even if it was not present initially. This can have relevant 
consequences on different baryogenesis models, which further stresses the importance of understanding the
interdependent cosmological evolution of magnetic fields and chiral asymmetry.
Finally, this description of the magnetic energy evolution and helicity creation could as well be relevant for baryogenesis models.

\acknowledgments
We thank Thomas Konstandin and Sacha Davidson for useful discussions that contributed to this work.
This work was supported by the ''Helmholtz Alliance for Astroparticle Physics (HAP)'' funded by the Initiative and Networking Fund of the Helmholtz Association and by the Deutsche Forschungsgemeinschaft (DFG) through the Collaborative Research Centre SFB 676 ''Particles, Strings and the Early Universe''.

\end{document}